\documentclass[preprint]{iacrtrans}
\usepackage{color}
\usepackage[utf8]{inputenc}

\usepackage{colortbl}
\usepackage{booktabs} 
\usepackage{tabularx}
\usepackage{rotating}
\usepackage{caption}

\usepackage{arydshln}
\usepackage{booktabs}
\usepackage{longtable}
\usepackage{rotating}
\usepackage{array}
\usepackage{multicol}
\usepackage{multirow}
\usepackage{graphicx}
\usepackage{array}

\usepackage{adjustbox}
\newcolumntype{L}[1]{>{\raggedright\let\newline\\\arraybackslash\hspace{0pt}}m{#1}}
\newcolumntype{C}[1]{>{\centering\let\newline\\\arraybackslash\hspace{0pt}}m{#1}}
\newcolumntype{R}[1]{>{\raggedleft\let\newline\\\arraybackslash\hspace{0pt}}m{#1}}

\usepackage{adjustbox}

\newcolumntype{T}{>{\tiny}l} 

\usepackage{tablefootnote}

\usepackage{pgf, tikz}
\usetikzlibrary{arrows, automata}
\usepackage{forest}
\usetikzlibrary{calc,positioning}
\usetikzlibrary{positioning}
\forestset{%
my forest/.style={%
    for tree={%
            node options={text width=3cm, align=center},
            l sep=1cm, 
            parent anchor=south, 
            child anchor=north,
            if n children=0{tier=word}{},
        edge path={%
            \noexpand\path [\forestoption{edge}] (!u.parent anchor) -- +(0,-15pt) -| (.child anchor)\forestoption{edge label};
        },
    }
}
}

\usepackage{lscape} 
\usepackage{amssymb}
\usepackage{lipsum}
\usepackage{subfig}
 \usepackage{soul}
\usepackage{blkarray}
\usepackage{bbding}
\pgfdeclarelayer{background}
\pgfsetlayers{background,main}
\usepackage{ctable} 
\newcolumntype{?}{!{\vrule width 1pt}}
\usepackage{smartdiagram}

\usetikzlibrary{arrows.meta,
                calc, chains,
                quotes,
                positioning,
                shapes.geometric}

\usepackage{forest,array}

\usetikzlibrary{shadows}
\author{Seyyedeh Atefeh Musavi\inst{1} \and Mahmoud Reza Hashemi\inst{1}}
\institute{School of Electrical and Computer Engineering, College of Engineering, University of Tehran, Tehran, Iran. \email{amusavi, rhashemi@ut.ac.ir}}
\title[\texttt{iacrtans} class documentation]{A Threat Modeling Framework for Evaluating Computing Platforms Against Architectural Attacks}

\begin{document}

\maketitle

\keywords[\publname, TCHES, LaTeX]{Architectural attacks \and System engineering \and Threat modeling \and Attack taxonomy \and Design structure matrix \and System security architecture \and Platform security \and Advanced persistent threat \and Security engineering \and Hardware-involved software attacks.}


\begin{abstract}
Architectural attacks are kinds of hardware-involved software attacks in which a software component misuse a privileged relationship with the hardware to by pass system protections, monitors, or forensic tools. These relationships are often not illegal and exist between system components by design.
 Hence, even a system with secure hardware and software components, can be architecturally vulnerable. Unfortunately, the existing threat modeling schemes are not applicable for modeling architectural attacks against computing platforms.
  This is mostly because the existing techniques rely on an abstract representation of a software (.e.g., Data Flow Diagram) as a primary requirement which is not available for a platform as a whole (considering both hardware and software elements).
 In this paper, we have discussed the necessity of a hardware-software architectural view to system threat modeling. Then, we have proposed Lamellae, a framework adapts threat modeling method to be applicable for untrusted platforms by a holistic approach. Lamellae involves system security architecture for abstract modeling of the platforms. Using the Design structure matrix analysis, Lamellae helps an end-user to identify possible attack vectors against a platform. The framework is a connection point of concepts from system engineering and software security domains. We have applied the framework on a multi-purpose computer with x86-64 architecture as a case-study to show the effectiveness of our framework.
\end{abstract}

\section{Introduction}
\label{intro}
Today it is hard to find any office, and industrial or medical center devoid of computing systems. Considering the serious consequences of cyber-attacks against such systems, preventive measures such as evaluating the security before deploying them in the field, is essential. 

We have witnessed a significant growth in low-level state-of-the-art attacks \cite{see}\cite{brossard2012hardware} \cite{pikit} \cite{dmakeylogger} against low-level components of computing platforms (including operating systems, VMM reference monitors, firmware, and even the hardware), in past two decades. This is while existing security solutions (e.g. host-based intrusion detection \cite{hids1}\cite{hids2}, and malware detection \cite{alazab}\cite{elhadi}, \cite{Shosha2012}) mostly concentrate on the security threats or vulnerability of application-level programs.

Hence a critical question is how the existing threats against these platforms are modeled to evaluate the security. The conventional answer includes software threat modeling efforts \cite{threatmodelbook} which is an accepted practice for predicting different attacks against a software element as early as possible in its life cycle. These efforts which can be done either by the developer or the end-user have rich literature and variety of tools to assess the security of an application against different user-level and kernel-level malicious attacks.

Sufficing to this software-only threat modeling procedure means that we have assumed other hardware components of the platform to be in the trusted computing base (TCB). Traditionally this has been a common assumption during security analysis. 


The assumption looks to be no more acceptable when considering recent state-of-the-art "\textit{Hardware-INvolved Software}" \cite{forristal} (HINS) attacks. HINS attacks may involve hardware at micro-architecture \cite{pikit} \cite{meltdown} \cite{spe} or architecture \cite{pikit}\cite{meltdown}\cite{spe} levels. 
Since the software cannot easily interact with the hardware, microarchitectural attacks include complex side-channel data leaks based on prioir knowledge of CPU microarchitecture. However, there are a broad range of more straight-forward attacks in which software misuse its legitimate access to hardware components of the system directly to apply any modification directly.

Architectural attack is a name for indicating such later multi-stage attacks misusing architectural features as noted by Zhang et al.\cite{zhang}. The origin for such attacks is some inconsistencies occur in cooperation of some components at HW-SW boundary. These include, as Letan et al. point, \textit{situation where every component seems to be working as expected, but their composition creates an attack path}\cite{freespec}. Hence, even a system with secure hardware and software components, can be architecturally vulnerable.


Although defedending against a single architectural attack does not need a new perception, analyzing the big picture may need. Due to increasing rate of such attacks in the current decade, we have witnessed a newfound research trend \cite{freespec}\cite{speccert} \cite{zhang} toward bringing up a holistic defensive approach to cover such threats. Hence, there have been no threat modeling framework available for such attacks which is the focus of this paper.


The inherent cross-layer nature of such attacks involve multiple components of a system in a complex scenario to conform a malicious mission. These cannot be easily identified by per-element applying of attack mnemonics. More important, almost all of these models rely on an abstract representation of their target as a primary requirement which is not available for a computing platform as a whole.

In fact, one needs to consider all of HW/SW components of the platform and their mutual relationships in order to be able to effectively evaluate the overall system security. It is however not a trivial task to consider all platform components with all their relations in a complex system. Considering the above noted points, we have to involve the system architecture (which is by definition concerned with components and their relationships) in the threat modeling process. System security architecture (SSA) is simply a view of overall system architecture from a security perspective \cite{SSA}. We believe that the SSA concept has not been attended enough in the system security literature.

In this paper we have proposed a new framework for threat modeling of a system with untrusted platform called "Lamellae". Accommodating current threat modeling practices for a platform, Lamellae  -which is a part of an ongoing research- has two main pillars: A system-centric approach to inter-component relationships through the platform and an abstract representation of the platform based on these relationships to achieve the platform's SSA.

\begin{itemize}
\item Adopting the existing threat modeling method in order to be applicable for predicting different rootkits and architecturall vulnerabilities in computing platforms (combination of software and hardware elements).  This includes proposing a DSM-based system security architecture representation, a taxonomy for inter-component relations misused by attackers, and set of matrix-based metrics to find potential attack vectors.
\item Applying the model on an X86-64 general purpose system as a case study and evaluating the study by some revealed Advanced Persistent Threat (APT) attacks.
\end{itemize}

In the remainder of the paper, after presenting a brief review of the related literature in section \ref{related}, we will see the details of Lamellae in section 
\ref{idea}. We will then elaborate on the use of the proposed framework to threat modeling of a sample system with x86-64 architecture as a case-study, in section \ref{case}. In section \ref{eval} we have evaluated the study by some APT attacks disclosed in recent years. Finally, sections \ref{discuss} and \ref{conclusion} will include discussion and conclusion, respectively.

\section{Related works}  
\label{related}

\subsection{Threat modeling}
Threat modeling is a traditional practice in software security (from 1977) with rich literature and tools \cite{threatmodelbook}. In recent years, there have been efforts toward adopting threat modeling for security assessment of more complex targets other than a single application. Cyber-physical systems \cite{dfd}\cite{tm-cps}\cite{almohri}, Autonoumous systems \cite{winsen2017threat}, cloud infrastructure\cite{tmcloud1}, embedded systems \cite{telehealth} are examples of fields borrowed the concept of threat modeling in their security evaluation process.

The most similar approach to the Lamellae could be seen in \cite{almohri} which has been proposed by Almohri et al. in the context of Medical Cyber-Physical Systems (MCPS). The authors have proposed their own threat model as well as a basic abstract architecture  for MCPS environments. However, like other researches exist, the platform-level threats mentioned in \cite{almohri} is too limited and lack a comprehensive view.

\subsection{Security architecture}
An abstract definition of security architecture defines it as a high level identification and describtion of components involved in providing system's security requirements \cite{lawlor2003survey}.
 Security architecture frameworks has often been studied in the context of enterprise architecture (EA). In this context security architecture describes "\textit{a structured inter-relationship between the technical and procedural solutions to support the long-term needs of the business}" \cite{sherwood}. SABSA \cite{sherwood2005enterprise} is a well-known example of such frameworks which are surveyed well in \cite{bahmani2010survey}.

These security architecture frameworks have been also extended to be used in specific technological domains. Chang et al. \cite{ccaf1} have considered security requirements of a cloud software (SaaS) in different phases of its life cycle (design, implementation, and test) to provide a data security framework for cloud environment. The security framework which is based on a multi-layered model called CCAF \cite{ccaf2} embodies data attack spectrum and also the technologies can be used to mitigate such attacks in a three-layer layout. The layers include access-control, identity management, and encryption.

 Trustcloud \cite{trustcloud} is another example proposed in the context of accountability in cloud computing domain. The framework considers five layers of abstraction including system, data, work-flow, regulations, and policies. Each layer is responsible for recording special logs which are required for providing accountability.

RAJAMKI \cite{cross-layer} proposed a design theory for resilient cyber-physical systems. The main component of the theory is a multi-layered reference architecture covering the human, software (cyber) and platform (physical) layers. Hahn et al. \cite{killchain} have proposed a security analysis framework for cyber-physical systems. The framework includes a layered architecture of such systems as well as a kill-chain model represents the security attacks. Mapping the kill-chain to the three-layer architecture (including cyber, control, and physical layers), enable the framework to analyze the security of the target system.

All mentioned frameworks - though considered the computing platform in their architectures- are such coarse-grained that they cannot be used in order to enumerate threats between sub-components of the platform. In fact, there is no prior framework for obtaining system security architecture (which is a concept with focus on a stand-alone computer system and excludes network connections and also applications running on the system [47]).

\input{flow}
\section{Lamellae framework}
\label{idea}
\subsection{ Motivation}
\label{motiv}
The main motivation for the Lamellae raises as the result of an empirical study of different architectural attacks (including but not limited to \cite{see}\cite{smmrootkit}\cite{acpi-rootkit}\cite{against-vt}\cite{dmakeylogger}) against computing platforms. The study shows that such attacks are leveraging a privileged relationship between components inside the platform which are often not illegal and exist between system's components by design. These kinds of attacks are not considered comprehensively neither in the existing attack taxonomies \cite{avoidit}\cite{cyberconflict} nor in threat modeling schemes. Other than different kernel/user level rootkits \cite{arsenal}, there are a broad range of attacks possible against lower components of the platform. Forristal has layed a foundation for classifying different types of such attacks in \cite{forristal}.

Threat modeling is a practice in security assessment of software application ecosystem. Due to growing number of threats against computing platform, we think that it is necessary to adapt such procedure in order to be applicable for platform's security evaluation. This would enable a security analyzer at the end-user party to predict possible attacks against an in-hand platform. Hence would be possible to answer questions like:
\begin{itemize}
\item Which platform is less vulnerable to architectural attacks? (comparing multiple platforms by their SSA)
\item Which platform components are better locations to host a specific security solution such that the solution cannot be bypassed?
\item Is there any potential architectural attack vector against the (SW/HW) TPM on this platform?
\item Which hardware/software extensions are better to avoid (when possible) from the security point of view?
\item Which hardware architecture is the most secure when using the specific operating system of the enterprise? 
\end{itemize}

\subsection{Challenges}
 Threat modeling of a software include three steps done recursively: Modeling the software via an abstract model like DFD, identifying threats (.e.g. using STRIDE\cite{torr2005demystifying}), and planning for mitigation of threats. In this practice either of threat modeling and mitigation planning are often performed in a per-element manner. For example whenever you have found an spoofing scenario, you can use authentication, signatures, tokens and similar choices for the spoofed resource to mitigate the threat.

 The first challenge of platform's threat modeling scheme is to find an abstract view of a whole platform covering all software, firmware, and hardware components. Data flow diagrams and state machines are the most prevalent decomposition model used for software application. Developers are asked to deliver the DFD of the target application in this step. This challenge can be seen in other adoptions of threat modeling practices like those use threat modeling in cyber-physical systems. As an example Yampolskiy et al.\cite{dfd} discuss the insufficiency of the DFD diagram elements for modeling such environments.

 In the context of untrusted computing platforms,  it is not a trivial task to achieve a DFD when having a whole computing platform in hand. In fact, DFDs are not sufficient as the platform security is highly concerned with control logic of the systems (Control Flow Diagram). This is because the platform is responsible to implement control flow redirections including API calls, interrupts, CPU mode changes, reconfiguration, and etc. This is while there has been no available procedure for achieving exact DFD/CFD of today's complex platforms for an end-user.

The second challenge is to find an appropriate guidance for identifying platform threats. Platform attacks are often multi-stage attacks involving multiple components of the system.
As was mentioned before, existing taxonomies \cite{avoidit}\cite{cyberconflict} don't cover all kinds of low-level platform attacks. Microsoft STRIDE also is not so helpful to guide the security analyzer to enumerate such cross-layer state-of-the-art attacks.

\subsection{Threat modeling}
\label{tm}
In this section we will accommodate the existing threat modeling procedure to be applicable for untrusted computing platforms. Figure \ref{flow} shows the steps required for our framework which are described in two following subsections.

\subsubsection{Matrix formation}
  At the first step and in order to handle the first above challenge, we have introduced a basic data-structure for indicating the System Security Architecture (SSA). The structure can leverage different
modeling perspectives
\footnote{Including structural, functional, behavioral, Rule-based, Object-based,
Communication, and Actor and role perspectives\cite{perspectives}.}.

Lamellae represent a platform by a simple structural modeling perspective in the form of a squared $N*N$ matrix called Design Structure Matrix (DSM). The structure is consistent with the popular layered view of the system architecture in which upper layers are less privileged. In addition, it has a rich supporting literature in the context of system engineering.

Hence, we suggest to use the SSA-based DSM instead of DFD which is achievable by information available to an end-user. SSA-based DSM can be seen as a more coarse-grained representation which is applicable for platforms with different software and hardware components.

Figure \ref{insight} shows the overall insight of the DSM matrix. Each system component has a specific row and column in the matrix. The granularity of these components is another decision point. Lamellae asks the user to consider standalone chips as a separated hardware component. For software and firmware components, the user could consider the independently developed code as separated components. We have suggested brief best practices for helping an end-user to find such components in four main classes including hardware, firmware, system software, and application-level components available in appendix section. For ease of use, we will call any hardware, firmware, or software component as a code execution unit.
\begin{sidewaysfigure}
\centering
\begin{adjustbox}{width=\textwidth}
\begin{forest} my forest,
  /tikz/every node/.append style={font=\normalsize  },
[  , for tree={s sep=1mm}  
    [, no edge, 
        [\textbf{Relation} ,no edge
           [\textbf{Attack family}, no edge
                [\textbf{attack examples}, no edge
                     [\textbf{ Subject and object of the relation in the sample attack}, no edge]
                ]
            ]
        ]
    ]
    [\textbf{System-level privileged relations}, no edge
        [Reflective
                [Attacks from same system layer
                    [thread injection\cite{exposed}
                         [user-level thread A has the privilege to inject into thread B]                    
                    ]
                ]
        ]
        [Logical
                [SW Man-in-middle                
					[Hooked system-call\cite{arsenal}
					      [The compromised OS is privileged enough to change system call pathes invoked by applications]
					]
                ]
        ]
        [Sequential
            [Foregone attacks
                   [Stoned Bootkit\cite{stonedbootkit}
                           [BIOS has sequential prevalence over OS MBR\footnote{Master Boot Record}]
                   ]
            ]
        ]
        [Control
            [Misuse of system-defined privileged control mechanisms
	            [SMM rootkit\cite{embleton}
                           [code running in SMM mode has control over the OS interrupt handler]	            
	            ]
            ]
        ]
        [Access
            [Misuse of accesses between system components
                [DMA Keylogger\cite{dmakeylogger}
                		[the GPU has DMA access to OS keyboard buffer]
                ]            
            ]
        ]
        [Protective
            [break security mechanism
                [Memory access control voilation\cite{ME2009}
                 	[BIOS should protect some parts of memory of Intel Management Engine]
                ]
            ]
        ]
        [Configuration
            [Change system configuration
                [SMI supression attack\cite{suppression}
						[SMI interrupt can be configured by north-bridge chip-set]                
                ]
            ]
        ]        
        [Physical
            [HW Man-in-middle
               [Malicious hard-disk controller\cite{mal-disk}
                   [Hard disk controller is physically privileged to the disk device to get a consistent with OS view.]               
               ]            
            ]
        ]
    ] 
]
\end{forest}
\end{adjustbox}
\caption{Proposed taxonomy of privileged relationships between components of the platform.}
\label{tax}
\end{sidewaysfigure}
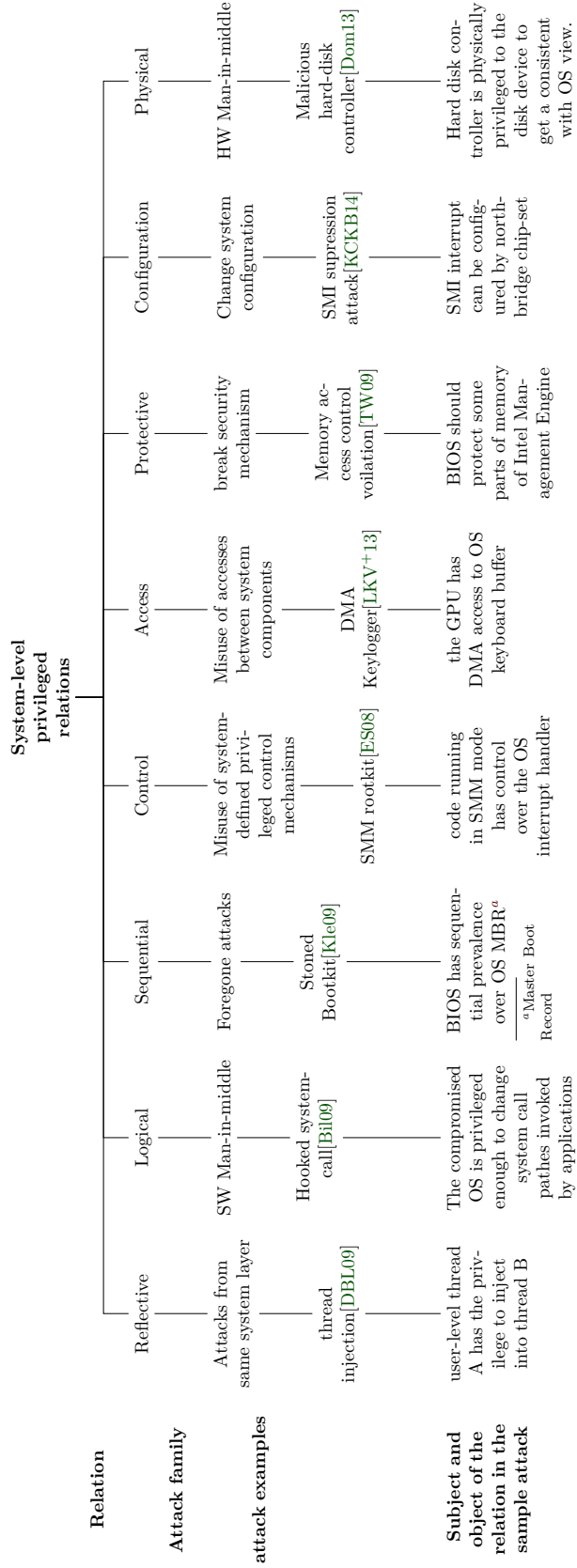

Code execution units are the basic block for the DSM rows and columns. The internal elements of the matrix should be filled by founded relations between different components. The $X_{i,j}$  element indicates that the $j$'th component has a kind of privileged relation to apply some modifications on the $i$'th element. The privileged relation has been recognized to be in the form of X category of the proposed taxonomy (Figure \ref{tax}) will be explained in continue .

Lamellae infer different types of relations based on possible attacks against platform.
We have performed an empirical study on the attacks against different platforms. In addition to a large number of related papers found manually from the Google Scholar website, we have tried to review related papers from the DBLP computer science bibliography database (over 3.3 million publications till 2016). To choose related papers from the whole archive, we have sufficed to about 220 unseen related titles . Considering each attack as a misuse of a privileged relation between the ordered pair of (attacker, attacked) components, we have found a generalization of such relations. Figure \ref{tax} shows our proposing taxonomy. Each kind of privileged relation can potentially cause different types of system-level attacks. The relations include:

1)	Physical privileged relation: Any platform consists set of hardware components which are arranged using some kinds of bus hierarchies. These hierarchies are often designed to improve the performance of the whole system \cite{bus-hierarchies}\cite{bus-hierarchies2}. However, from the security point of view the physical medium can be a source of physical man-in-the-middle attack. In other words, the attacker resides in higher levels of bus, the hierarchy is physically privileged over lower components and can modify the data owned by the component.

2)	Logical privileged relation: Systems are designed so that they can provide the desired abstraction for software developers. This is often implemented such that any code has logical twins run in lower levels of the system. For example, when a Java application calls for a system service, there may be codes in the interpreter as well as the operating system which run in turn to execute the desired service. These twins which translate the more abstract commands coming from their higher levels, can be implemented maliciously.

3)	Sequential relation: Executable codes may form a partially ordered set based on the time they run after the system turns on. This ordered relation may be valid continuously, or temporary. The component which its code runs sooner, has the ability to apply its desired changes without interception of the next components.

4)	Configuration relation: Components which can change some system's configurations can cause attacks by decreasing security level during execution of other components. While this kind of relations makes the system security analysis sophisticated, it is itself a result of complexities and heterogeneity of the system design.

5)	Control privileged relation: A system often includes some legal facilities to control, manage or monitor some components. These control power often can be lever-aged by malware developers for attacking the con-trolled component. In other word a counter attack can occurs by the malicious controller itself.

6)	Reflective privileged relation: Any code can be vulnerable to attacks sourced by code running at same privilege level with similar access rights. For example, kernel drivers can modify the OS kernel components as they both run at CPU kernel-mode.

7)	Access privileged relation: The more coupling between system components, the more mechanisms of interaction and access permission are needed to be defined in-between. The system may include some components with no negligible levels of coupling as the result of weak initial design or when providing extendability. These coupling and permitted access rights between components can be misused in an attack scenario.

8)	Protective relation: A component may be protected by the use of access control or integrity checking mechanisms. Since the completeness of such protections could not be proven formally in all situations, we consider a probability for the mechanism to be bypassed.

 Hence we have solved the second mentioned challenge by proposing our octet architectural relationships which we believe can better reveal malicious activities through a given platform.

One of the advantages of modeling the security relations between the platform components based on these privileged accesses is that they can cover potential attacks which are still not implemented for a specific platform. This is because the taxonomy is achieved by generalizing the inherent nature of different kinds of platform attacks. However, it should be noted that there would be no provable guarantee for these classes to cover any new attack with new inherent characteristics. In fact,
if we can imagine a new type of attack which exploit a privilege relationship which does not exist in our taxonomy, it should be used to update the taxonomy. After the update, all of different attacks with this new characteristic are covered by Lamellae.


The order of rows (and their corresponding columns) is important in Lamellae. The matrix should be arranged such that each row has some kind of privileged relation over all its higher rows. This comes from the initial insight of the framework in which the computing system is seen as a layered system in which lower layers are more privileged. In the DSM literature, there are few rules to be applied on a raw DSM in order to ensure that the units are chosen and arranged in the right order \cite{dsm-methods}:
\begin{itemize}
\item Sequencing rule: Similar to sequencing procedure for DSMs, we move rows and columns such that the filled cells below the diagonal go to top of the diagonal or if not possible as near as possible to it. Hence entities with empty columns and rows (excluding elements on diagonal) should be shifted to the most right and left columns respectively.
\item Integration rule: Rows with identical marked columns will composed into a single layer. The integrated units are called circuits in the context of DSM processing. There have been multiple proposals for finding circuits inside a matrix \cite{dsm-methods}\cite{circuits}.
\item Precedence rule: Row A will placed higher than row B if its filled cells are subset of row B.
\end{itemize}
\begin{figure*}[t]%
    \centering
    \subfloat[A, B are coupled activities]{{\includegraphics[width=3.2cm]{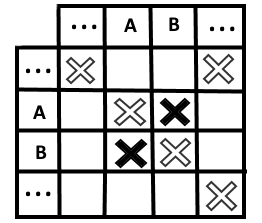} }    \label{dsm1}
}%
    \qquad
    \subfloat[B has a removed tear,]{{\includegraphics[width=3.2cm]{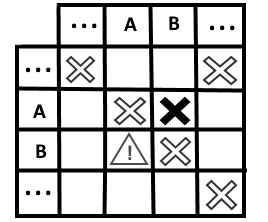} }    \label{dsm2}}%
    \qquad
    \subfloat[B and C are conditional activities depends to A ]{{\includegraphics[width=3.2cm]{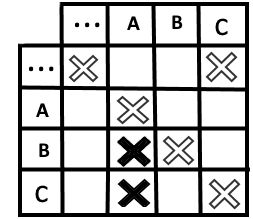} }    \label{dsm3}}%

    \caption{\footnotesize{Different analysis concepts in the context of DSM analysis}}%
    \label{dsm}%
\end{figure*}
\subsubsection{Matrix analysis}
\label{analysis}
After applying the rules, the final matrix which describes the security architecture of the target system is ready. Hence we have achieved an architectural model of the platform.
In the next step we need to find threats for a specific given platform. In other word, we should analyze the in-hand matrix to see "how can things go wrong"\cite{threatmodelbook}?

If the system shows a fully layered scheme (upper-triangular matrix), it is clear that the platform would be more easy to be secured. This is because a probable security solution placed in each layer can be attacked only by its lower layers. So most kinds of STRIDE scenarios can be prevented by fixing monitors in the more privileged levels. By placing a powerful monitor at the lowest trusted component one would theoretically be able to monitor all components of the platform securely.

However, in most cases the system has some other relations below the diagonal. Hence, possible attack vectors against protection mechanisms embedded in any component should be analyzed more accurately. DSM analysis include few metrics for reflecting such violation of layered design which include \ref{dsm}:

1)	Coupled sub-systems: Two entities are coupled when there is an inter-dependency between them and either have privileged relation over another. The higher number of coupled execution units, the more probability of security solutions to be ineffective for the target system. This is because the solutions which are designed to protect the less privileged component, could themselves be attacked by it. One of the main goals of DSM analysis is to find out such coupled sub-systems. Figure \ref{dsm1} shows a DSM containing such units.

2)	Removed tears: tears can be seen as a coupling which has been tried to be removed at design time, however it has not been removed totally. At design time analysis, tearing is a process in which some feedback marks (tears) are removed from the matrix by adding some assumptions about the system \cite{dsm-methods} (figure \ref{dsm2}). They often are implemented by adding a limitation satisfying the made assumptions in the platform. Depending on how much robust the limit protection has been implemented, the tear can be seen as a weak point of the platform.

3)	Conditional executions: Conditional units can be seen when a component has privileged relations to multiple other components (figure \ref{dsm3}). From the security point of view, the base units -unit A in the example- should be designed more robustly or be defined inside the system TCB. Conditional units below the platform's DSM diagonal could be a more severe security threat. This is because such components could violate the layered design and by passing the layered security view in multiple points.


In continue, we will explain our DSM-based attack vector identification which include a respective matrix-based metric for each of possible architectural attacks\footnote{It is worthy to note that our metrics only indicate potential attack scenario and the possibilty of the attack in real world depends on some other factors including more fine-grained focus on details, attacker skills, and etc.} (Forristal's taxonomy\cite{forristal}). Table \ref{metrics} show these metrics which are explained in continue:

1) Inappropriate General Access to Hardware\footnote{ "\textit{a higher-privileged software layer (i.e. OS or hypervisor) incorrectly grants or proxies general hardware access to a lower-privileged software layer}"\cite{forristal}.}: These attacks are often called \textit{confused deputy attack} in the mobile security literature. 
 From the DSM point of view, any occupied cell at the rows belonging to the system software components below the diagonal can indicate such attack vectors. We have also considered  "Interference with Hardware Privilege Access Enforcement"\cite{forristal} attacks as a special case of inappropriate general access.

2) Hardware Reflected Injection\footnote{\textit{The attacking application originates particular malicious
data, and that data traverses through the higher-privileged software stack layers
and into the hardware for storage; later, a privileged software layer
receives/retrieves the malicious data from the hardware, and immediately operates upon, interprets, or
otherwise uses the malicious data in an insecure manner leading to a security vulnerability.}}:
 We will consider the equivalent node-link diagram of the DSM occupied cells to find these kinds of injections.
Existing relation between two components is shown via a directed like between two nodes in such graphs
\footnote{ More details on such graphs can be read in \cite{browning2016design}.}.
Hence such attack scenarios can be detected by finding all anti-clock-wised triangular walks start and end through the upper-diagonal part of the matrix in which the middle point of the walk is at the below of the diagonal.

3) Access By a Parallel Executing Entity: Here we have a row with multiple occupied cells. 
Handling such parallel executions in one of the two main tasks of the operating system. There are also some of micro-architectural mechanisms to handle them at hardware-level. However the existing vulnerabilities shows that sometimes the mechanisms fail to handle specific conditions. 
Hence any horizontal line between occupied cells (excluding sequential/logical relations) in the rows belong to computing resources can be a potential parallel executing attack vector and should be monitored more accurately at run-time.

4) External Control of a Hardware Device: An external software component can control a device in this attacks. Such a software component may enter the system through the network stack, or by other interfaces with physical access to the system (.e.g. Firewire). Any $X_{i,j}$ in which $i$ regards to hardware device row, and $j$ is a column from  external hardware interfaces/devices should be checked for detecting such potential attacks.

5) Unexpected Consequences of Specific Hardware Function\footnote{These attacks may occur in condition in which 
an access to a hardware is granted to a platform's component with the \textit{" presumption that only particular hardware functionality is exposed, and access to that specific hardware functionality is not believed to harbor any security concerns"}\cite{forristal}.}:  It is a fact that Lamellae is designed to prohibit such unexpected consequences. Here we expect the analyzer who fill the matrix to accurately read between lines of hardware manuals to reflect all possible privileges and access rights in the matrix.

\begin{table*}[t]
   \caption{Analysis of hardware-involved software attacks by Lamellae} 
   \label{metrics}
   \footnotesize 
   \centering 
   
   \begin{tabular}{ p{2.5cm} p{2.5cm} p{3cm} p{4cm} } 
   \toprule[\heavyrulewidth]\toprule[\heavyrulewidth]
 
   \textbf{Attack\cite{forristal}}  &\textbf{example} & \textbf{Metric} & \textbf{definition}\\ 
   \midrule  
   Inappropriate General Access to Hardware & CVE-2010-1592 &Checking specific cells&occupied cell at the rows belonging to system software/firmware components below the diagonal\\\hline
   Hardware Reflected Injection& CVE-2010-4530 & quadrangular walks & anti-clock-wised triangular walks start and end through the upper-diagonal part of the matrix\\\hline
   Access By a Parallel Executing Entity & CVE-2010-0306&horizontal line& horizontal line in the rows including computing resources\\\hline
External Control of a Hardware Device &CVE-2011-3215&Checking specific cells&$X_{i,j}$ in which $i$ regards to hardware device row, and $j$ is a column from  external hardware interfaces/devices\\
   \bottomrule[\heavyrulewidth] 
   \end{tabular}
\end{table*}

\section{case-study}
\label{case}
 In this section we will see how the Lamellae can be applied on a system with Intel x86-64 architecture (corei7 Sandy Bridge). This would enable us to evaluate how much the framework can cover real rootkits in practice. The motherboard of the system has a traditional architecture in which south-bridge and north-bridge are separated. Our first source of study should be the Intel processor's manual. We have found initial code execution units could be classified into 17 main classes.
In three below subsections we will review relationships among code execution units classified into hardware, firmware, system software, and application levels. For each unit two kinds of possible attacks should be mentioned through the descriptions: Attacks can be sourced by the unit, and also attacks perform by other units targeting the specific unit. The relationships will be reflected in a DSM matrix (Table \ref{INAM}). At the end, we will analyze the final DSM matrix from the security point of view.

\subsection{Hardware execution units}
 Hardware execution units consist of CPU and any device connected to the Central processing unit on the mother-board. We have excluded other electronic devices. Hardware execution units include: 
 
- CPU hard-wired code/state: From the security point of view, the CPU includes mainly hardwired codes as well as configuration registers which if include malicious functionality (like \cite{mal-cpu}\cite{mal-cpu2}) can affect all execution units run on the CPU. Thus the CPU has logical privileged relation to all other units. As an example the CPU can be configured to use cache instead of SMRAM in order to help the attacker to implement a SMM rootkit without requiring modifying the SMI handler in the SMRAM \cite{cachep}. It also can disable required security technology like interrupt remapping protection \cite{against-vt} which facilitates some attacks against different components including BIOS, OS and etc. CPU states however could be set/modified by the platform firmware as well as the system software. Hence here these execution units have sequential privileged relationship with the CPU state. For example, recently advanced rootkit has shown the ability of modifying the system memory layout by installing a kernel module \cite{see} (control privileged relationship).

- North bridge: The next execution unit corresponds to the north bridge, which physically is privileged since it can mediate the data that comes from all other devices before being received by the CPU.

- North bridge devices: Typically, main memory and graphic hardware are directly connected to the north bridge. The physical position empowers either rootkits \cite{mguard} and also security monitor developers to access the main memory (and thus software components running on the system) by DMA. Recently, stealth GPU-assisted rootkits has been proposed \cite{gpu-assisted}, which can be fully invisible in the main memory. Another often ignored privilege relationship is the configuration relation between the devices and the BIOS code which is described by SIPI attack. In \cite{against-vt} it has been shown that a using a message signaled interrupt (MSI), a device can force the CPU to run the custom start-up boot code and so by pass the BIOS. In the same time there are many kinds of firmware and software components have access privilege to the devices.

- Out-of-band management component: Out-of-band (OOB) management is a technology enabling control and management of an Intel platform independent of the CPU. Current versions of the technology for intel motherboards contain Intel Management Engine (ME) which is a small computer able to access the whole system's memory (excluding some system-reserved regions), the CPU, and the power management of the two \cite{MEbook}. Hence it can intercept most of system components by a counter attack (It can inject operations to the CPU, use its dedicated DMA hardware to read and write to the memory of OS and applications) \cite{MEbook}.

 There are discussions about how the technology can be used to implement a rootkit targeting the CPU, ROM and also RAM of the system \cite{ME2011}\cite{ME2013}\cite{dagger}, thus it would be a threat for all levels of the system. This special position of ME engine, has made it interesting for malware developers such that "ring -3 rootkits" \cite{ME2009} referred to rootkits residing in it. ME resides in the north-bridge chip set, thus the chip itself has a kind of physical privileged relationship with it. At the first glance it may seem that the engine is fully independent of the system, however there are two points it relies on the system BIOS. The ME boot loader loads its code from the ME ROM and BIOS has temporary sequential relationship with it when it borrows from the system RAM . Consequently, these memory pages are potentially accessible by devices which can access the RAM. After some attacks shown the possibility of code injection to these BIOS reserved pages \cite{ME2009}, a new page integrity schema has been proposed to ME engine. Hence we consider a probability for the scheme to be bypassed as a result of limitations it has due to performance considerations.

- Southbridge: Southbridge is another chip which connects slower devices (like PCI, USB, IDE devices, and ROM memories) to the Northbridge. Thus peripheral devices as well as BIOS chip are under the physical privileged accesses of the bridge.

-Southbridge devices: These devices may include the DMA engines have been used to access the system main memory as the result of access relationships \cite{dma-windows}\cite{dma-pci}\cite{lophi}. Here again there is the possibility for SIPI attack mentioned before by the configuration relation of devices with the BIOS code.
 System software (OS/hypervisor) is sometimes is allowed to flash the firmware stored in hardware devices to update their code which is shown by access relationship inside the matrix. 

\subsection{Firmware execution units}

-Micro-code: One may think that the CPU codes are all hard-wired, however today's CISC processors mostly use a microcode-based architecture in order to translate some complex instructions. It is evident that the microcode modification could modify all levels of code executions, which run by the processor, thus it provides logical privilege relationship of the CPU to other codes. For Intel X86, the micro-code which is stored in an internal ROM chip, is also updatable at boot-time by the system BIOS. Hence, the BIOS code has a temporary sequential privileged relationship with the system microcode.

- System BIOS: System BIOS code is physically stored in flash chips connected to the south-bridge, and has an important role in system boot-up. System BIOS is responsible for hardware self-testing, initialization of the hardware devices, and the OS loading. It is the first software code run on the system, hence it can be a source of attacks in multiple levels. The BIOS can intercept configuration of devices, handler for SMI Interrupt \cite{lighteater}\cite{biossmm}, Measured Launched Environment (MLE), or the system software as it has a sequential prevalence to all of these components.

- UEFI: UEFI is a newer CPU-independent firmware interface architecture aims to replace the legacy BIOS. It can be seen as a small OS (C program) running on protected mode before the main OS. Hence attack vectors for UEFI is similar to the BIOS. Despite, UEFI also can sit on top of the legacy BIOS.

- Option-ROMs: Option-ROMs which are defined by PCI specifications, which are firmware components located on each device separately. They are loaded by the system BIOS at the boot time for providing additional functionality or replaced the existing BIOS service required by the device.

\begin{landscape}
\begingroup
\setlength{\tabcolsep}{8pt} 
\renewcommand{\arraystretch}{1.4} 
{\scriptsize
\begin{table}[t]
\centering
\caption{DSM matrix of the studied case. The symbols used in the table include $P$ for physical, $L$ for logical, $S$ for sequential, $C$ for configuration, $T$ for control, $R$ for reflective, $A$ for access, and * for protective relationships. }
\begin{tabular}{|p{0.3cm}|p{3cm}|p{0.25cm}|p{0.25cm}|p{0.25cm}|p{0.25cm}|p{0.25cm}|p{0.25cm}|p{0.25cm}|p{0.25cm}|p{0.25cm}|p{0.25cm}|p{0.25cm}|p{0.25cm}|p{0.25cm}|p{0.25cm}|p{0.25cm}|p{0.25cm}|p{0.25cm}|p{0.25cm}|p{0.25cm}|}
\cline{3-19}
\multicolumn{2}{c|}{} & \rotatebox{270}{application} & \rotatebox{270}{Run-time env} & \rotatebox{270}{OS libs/services}& \rotatebox{270}{Operating system} & \rotatebox{270}{Hypervisor} & \rotatebox{270}{MLE}& \rotatebox{270}{SMI code} & \rotatebox{270}{option-ROMs}& \rotatebox{270}{UEFI}& \rotatebox{270}{BIOS}&  \rotatebox{270}{s-bridge dev)}& \rotatebox{270}{Southbridge}&\rotatebox{270}{n-bridge dev}&\rotatebox{270}{OOB management}& \rotatebox{270}{North-bridge}& \rotatebox{270}{microcodes} & \rotatebox{270}{CPU state}  \\
\hline
\multirow{ 5}{*}{\rotatebox{90}{\textbf{APP}}} 
&Application
 & $R$\cellcolor[gray]{.8}
 & $L$ 
 & $T$ 
 & $L$ 
 & $L$
 & $T$ 
 & $T$ 
 &$S$
 & $S$
 & $S$ 
 & $A$
 & $A$
 & $A$
  & $T$
 & $P$ 
 & $LC$
 & $LC$
\\\cline{2-19}
&Run-time env   
 & $A$
 & $R$\cellcolor[gray]{.8}
 & $T$ 
 & $L$ 
 & $L$ 
 & $T$ 
 & $T$ 
 &$S$
 & $S$
 & $S$ 
 & $A$
 & $A$
 & $A$
  & $T$
 & $P$
 & $LC$
 & $LC$
\\\cline{2-19}
& OS libs/services   
 &$A$
 & $A$  
 & $R$\cellcolor[gray]{.8}
 & $L$
 & $L$ 
 & $T$ 
 & $T$  
 &$S$
 & $S$
 & $S$ 
 & $A$ 
 & $A$
 & $A$
  & $T$
 & $P$
 & $LC$
 & $LC$
 \\
 \hline
\multirow{ 4}{*}{\rotatebox{90}{\textbf{Sys SW}  }}
& Operating system   
 & $*$
 & $*$
 & 
 & $R$\cellcolor[gray]{.8}
 & $L$ 
 & $T$  
 & $T$  
 &$S$
 & $S$
 & $S$ 
 & $A$
 & $A$
 & $A$ 
  & $T$
 & $P$
 & $LC$
 & $LC$
\\\cline{2-19}
& Hypervisor  
 & $*$
 & $*$     
 &
 &
 & $R$ \cellcolor[gray]{.8}
 & $T$
 & $T$ 
 &$S$
 & $S$
 & $S$ 
 & $A$
 & $A$
 & $A$ 
  & $T$
 & $P$
 & $LC$
 & $LC$
\\\cline{2-19}
& MLE 
 & &  &  &   &  
 & $R$\cellcolor[gray]{.8}
 & $T$ 
 &$S$
 &$S$ 
 & $S$ 
 &
 & $P$
 & 
  & $T$
 & $P$
 & $LC$
 & $LC$
 \\
\hline
\multirow{ 5}{*}{\rotatebox{90}{\textbf{Firmware} }}
&SMI code & &  &    &  && 
 & $R$\cellcolor[gray]{.8}
 &$S$
 & $S$ 
 & $S$
 &
 & $P$
 & 
  & $T$
 & $P$ 
 & $LC$
 & $LC$ 
\\\cline{2-19}
&Option-ROMs  && &  & & &
 & $T$ 
 & $R$\cellcolor[gray]{.8}
 & $S$ 
 & $S$
 &
 & $P$
 & 
  & $T$
 & $P$
 & $L$
 & $L$
 \\
\cline{2-19}
&UEFI  & &  &
 &$T$
 &  $T$
 & 
 & $T$
 & 
  & $R$\cellcolor[gray]{.8}
 & $S$
 & $C$
 & $P$
 & $C$ 
  & $T$
 & $P$
 & $L$
 & $L$
\\\cline{2-19}
&BIOS  & &  &  
  &$T$
  &$T$
  & 
 &$T$ 
 &  & 
 & $R$\cellcolor[gray]{.8}
 & $C$
 & $P$
 & $C$ 
  & $T$
 & $P$
 & $L$
 & $L$
 \\
\hline
\multirow{ 7}{*}{\rotatebox{90}{\textbf{Hardware} }}
 & S-bridge dev  &    & & &$A$ &$A$&  
 & 
 &
 & 
 &
 & $R$\cellcolor[gray]{.8}
 & $P$
 & 
  & 
 & $P$
 & $L$
 & $L$
\\\cline{2-19}
&Southbridge& &   && & & &  & 
 &  
 & 
 &
 & $R$\cellcolor[gray]{.8} 
 & 
 &   
 &$P$
 & $L$
 & $L$
 \\
\cline{2-19}
 &N-bridge dev  &    &  
 & 
 &$A$
 &$A$ 
 & $A$ 
 & $A$ 
 & $A$ &$A$  &$A$ & &
 & $R$\cellcolor[gray]{.8}
 &
 & $P$
 & $L$
 & $L$
\\
\cline{2-19}
&OOB management  &  &  & &$*$  &$*$ 
&
 &$*$
 & $*$ 
 & $*$
 & $*$
 & $*$
 & 
 & $*$
 & $R$\cellcolor[gray]{.8}
 & $P$
 & 
 & $*$
 \\
 \cline{2-19}
&North-bridge & &  &   &
 &
 &
 &
 & 
 &  
 & 
 &  
 &  
 & 
 & 
 & $R$\cellcolor[gray]{.8}
 & $L$
 & $L$
\\\cline{2-19}
&micro-codes   &  &   & &  && &  & &
 &$S*$ 
 & 
 &
 & 
 &$T$
 &   
 &$R$ \cellcolor[gray]{.8}
 & $L$ 
 \\
\cline{2-19}
&CPU hard-wired code/state   
 &*  
 &* 
 &
 & $T$ 
 & $T$
 &
 &
 & 
 & $S$
 &$S$ 
 &  
 &
 &
 &$T$ 
 &   
 &
 & $R$ \cellcolor[gray]{.8}
 \\
\hline

\end{tabular}
\label{INAM}
\end{table}
}
\endgroup
\end{landscape}
 In practice, an option ROM can have unrestricted access to the system when running \cite{optionROM}. These codes have full access to the system and have sequential privileged relationship with other components loaded into the memory later.

- SMM code: X86 CPUs have a system management operating mode (SMM mode) to handle the platform specific functionalities in a separated address space (SMRAM) which is transparent to the system software. In recent years the mode has been focused by either attackers and also security productions \cite{stm-userguid}. The system management interrupt (SMI) handler is primarily stored in BIOS and is loaded to the SMARAM at boot-time (temporary sequential relationship). It is also " neither visible, nor maskable" \cite{stm-userguid} by the system software as working in a separated processor mode. SMI interrupt can be generated either by hardware (signaling the SMI pin on the processor) or by the software (Message Signaled INT). From the security point of view, SMM mode has become a great source of counter attacks as it can access the memory entirely (control relation). It can be used as a vector to attack option-ROMs of peripheral devices (like a malicious IOCHECK \cite{iocheck}), hypervisor, operating system, or any kind of user-level applications. It also has been used to attack the MLE \cite{invisible}.

- Measured launched environment: Intel x86 TXT is a hardware extension which aims to provide a trusted plat-form by the help of a TPM module. By using the extension, the processor would be able to verify the system software integrity and the platform configuration. If the platform passes the verification, then a pre-defined known state of the platform would be applied to the system. The MLE could be used at boot time or be a late-launch session inside the operating system. Once the MLE is launched, it can run an uninterrupted environment for a binary. Although the facility is generally used by white-hats \cite{lala}\cite{flicker}, it has been shown \cite{cloak-comp} that it can be used also by black-hats to run an anti-analysis malware with no interruption or possibility of analysis. Though, this control relation could be a threat itself for higher level units.

\subsection{Software execution units}
- Hypervisor: In X86 architecture, hypervisor can work either by hardware or para-virtualization. In either of cases the VMM has logical privileged relationship with the units run later inside the virtualized environment. This relation has made it possible to implement stealth malware attacks \cite{v-black}. The logical dependency is especially interesting for the attacker who want to hijack the current operating system by installing a hypervisor below the OS  (blue-pilling) \cite{bluepill}. Though the SMM code has been proposed as an integrity checker for the hypervisor \cite{hypercheck}\cite{hypersentry}, it can also be used as a counter attack to the hypervisor because of the existing control relation.

- Operating system: The OS inherently has a complete control of the execution units run inside the box as the result of logical relation to other user-level unit. In addition, currently ACPI power management is supported by most of the operating systems will introduce new set of threats from the security view. The ACPI scripts run by the AML interpreter of the operating systems are expected to be used for power control. However, they can disable security functionalities implemented by a separated hardware devices or by interrupt remapping \cite{against-vt}. They also can run other arbitrary code as a cross-platform rootkit \cite{acpi-rootkit}. Thus devices have access relationship to the OS. The scripts in turn can be re-written at boot-time by the BIOS (sequential relation).

- User-level OS libraries: Unlike traditional monolithic operating systems, there is a growing trend toward minimizing the kernel and to outsource some OS services to the user-level execution units. The Libraries have logical privileged relationship to ordinary application programs.

- Run-time environments: Sand-boxes, containers, and run-time frameworks are examples of runtime environments manage the code running inside. Thus these are very similar to ordinary applications from the security view. There are special types of rootkits called managed code rootkits \cite{mcr}, which aim to change the behavior of the environment. Since the runtime environments executes at user-level, such rootkits are in fact user-level rootkits that look like kernel-level ones from inside of the environment (access relation).

- User-level applications: User-level application (either interpreted or native) has access relationship with the user-level OS components. In fact, applications have same privilege level to the user-level OS libraries, hence they can use the legitimate APIs or OS facilities in order to penetrate into the OS by a user-level rootkit \cite{exposed}. Although installing malicious kernel modules in order to penetrate into the kernel requires administrator access, however this is a common procedure during application installation. Hence this is a weak protective relation that can be easily bypassed by kernel-level rootkit \cite{kernelrootkit1}\cite{kernelrootkit2}.
\subsection{DSM processing and analysis}
\label{sec-ana}
After raw insertion of mentioned privileged relations into the matrix, we have applied the rules of sequencing and integration discussed in section \ref{tm}. As a result, we will see different parts of the operating system (including kernel and drivers) made a single circuit. Table \ref{INAM} shows the final DSM matrix. As the table suggests there are multiple execution units (like native and interpreted applications, OS components, and etc.) which have been integrated into a same layer.

The upper triangular part of matrix enables the analyzer to find a primary layered view of the target system. For example, since the option-ROMs are placed at the bottom of the OS, the analyzer would note that it would not be sufficient to protect the application's memory after OS has been load-ed. Instead, it is required to ensure that the OS has not been compromised by DMA attacks (even though the initial OS and the boot process are trusted).

The lower triangular part of matrix, reflect the design violations of this initial layered view. These violations result in more complex vulnerabilities which often are not easy to handle. Below, we will analyze the matrix by the approach proposed before to see the design vulnerabilities as the result of elements below the diagonal.

First of all, we analyze platform couplings (figure 3a). We have found the following coupled units which can violate the pure layered representation inside the matrix:

1) Update couplings: There are multiple update couplings in the system where the higher unit is able to update the lower one. One of the most threatening one is BIOS-CPU coupling. The platform firmware (BIOS/UEFI) is executed under the control of the CPU hardware. However, in X86 the system firmware in turn has sequential privileged relation to the CPU microcode. In fact, the system has been designed in a way that the firmware updates the microcode and initializes the CPU registers. This will make it hard to make a chain of trust based on the correct functionality of the CPU and the CPU cannot check the integrity of the firmware. This is be-cause the CPU itself may have been attacked by the BIOS. Although there is no implemented microcode attack against Intel microcode updates, the coupling has been exploited in an AMD processor before \cite{microcode}. Other than microcode update, there are other similar cases including updating BIOS/UEFI by the system software, updating the OS by the application and etc.

2) OS-application coupling: As we expect the OS has logical privileged relationship with the applications. However, the ability to install a driver in x86 architecture by admin privilege, enable applications to penetrate into the kernel. This will violate many security assumptions about the user-level apps. This weak protection (alert dialogue to be verified by the user) can be bypassed by social engineering techniques to convince the user to install malicious drivers.

3) ME-platform coupling: As was noted, the management engine of the Intel has its own memory; however, it also uses a borrowed memory from the system's RAM. Hence considering the threat of access control to be bypassed (has been shown to be possible before \cite{ME2009}), all software units which are able to access the main memory can modify this borrowed section.

Another point which was noted in section \ref{idea} is the removed tears (figure \ref{dsm2}). The final matrix shows that a coupling can be seen in between the Intel MLE and the SMI code runs at system management mode. MLE is designed in a way that it does not rely on any code which has not been measured before. However, while the MLE late launch session is sequentially run after BIOS, the SMI handler (as a part of BIOS (can run then inside the MLE independently. Evidently this can neutralize the measured environment of the MLE. The problem has been raised because the hardware manufacturer assumes the handler to be trusted and benign. Late after about a decade SMI attacks have been revealed, Intel is trying to tear the coupling and recently released a specification for a monitor called SMI Transition Monitor (STM) \cite{stm-userguid}, in order to control the assumption and to monitor SMI handler when an MLE is launched.

Conditional units are the third item should be considered (figure \ref{dsm3}). As an example by considering the SMM column below the diagonal, it can be seen how adding a new mode can violate the layered architecture of the system. 

\begin{figure}
  \centering
    \includegraphics[width=0.42\textwidth]{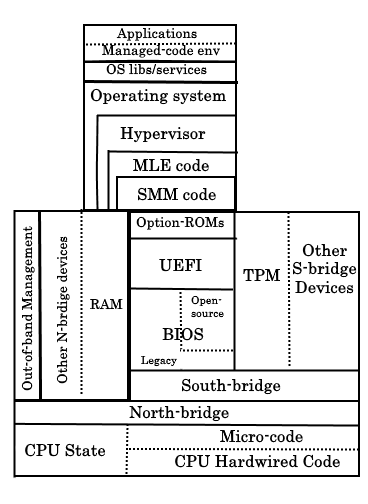}
      \caption{Graphical representation of architecture embedded in table \ref{INAM}}%
      \label{layered}
\end{figure}

To better understand the system, it may be useful to convert the matrix to a graphical representation. A fully layered system (fully upper triangular DSM) would be easy to represent graphically. Each row in DSM matrix would form a layer on top of its lower row. Layers which are not necessary for a system to work, they can be represented in an L form (for example a system may or may not have a VMM). Rows which include some layered-view violations should be placed beside each other as the result of couplings exist among them. Graphical view in figure \ref{layered} has no more information than the matrix, but maybe it can better show how much the primary fully layered imagination about the system was true in practice.

As it was noted in section \ref{tm}, the analysis can be continued in a recursive manner if lower granularity is desired.
 As an example, let's consider a threats at UEFI level. After analyzing the security of a code resides in UEFI position in the matrix, analysis can be continued to focus on twelve stages of boot sequence in the mentioned platform.


Finally by the platform's DSM, one can think about the questions mentioned in section \ref{motiv}.
As an example a critical question is about which platform components are better locations to host security solutions such that the solution can not be bypassed? The matrix shows that - even if we trust the hardware components- none of software or firmware components are appropriate locations for an ideal security solution. This is because there are layered violation in all rows of such components. When discussing hardware, - other than the CPU- the north-bridge looks to be an ideal place in our case. In fact, if one can assume the CPU in the TCB, north-bridge chipset is the most privileged place with no below diagonal tear. This maybe the reason why Intel has chosen the chipset for placing AMT engines.

\section{EVALUATION}
\label{eval}
In this section we will evaluate the Lamellae based on the case-study mentioned in section \ref{case}. We have achieved the DSM matrix of the target system which reflects the security architecture of the platform.
 A serious challenge for evaluating any APT-related research is the lack of available instances occurred in the wild. Low-level platform attacks are infrequent among huge amount of malware detected every day. Hence it is not trivial to evaluate the Lamaellae based on these kind of targeted attacks.

\begin{sidewaysfigure}
   \caption{Platform attacks against x86-64 occurred in the wild which are shown to be prospected by Lamellae} 
   \label{tableeval}
   \centering 
   \begin{tabular}{p{2.3cm} p{0.7cm} p{0.7cm} p{4.5cm} p{2cm} p{6cm} p{3cm}} 
   \toprule[\heavyrulewidth]\toprule[\heavyrulewidth]
   \textbf{Rootkit} & \tiny{\textbf{Occured}} & \tiny{\textbf{Detected}} 
    & \textbf{Details} &\tiny{\textbf{Misused privilege}} & \textbf{attack type} & \textbf{Cells involved} \\ 
   Alureon \cite{alureon}& 2010&2010   &  Changes LoadIntegrityCheckPolicy boot option & Configuration & HW Reflected Injection & \raisebox{-.6\height}{\includegraphics[scale=0.2]{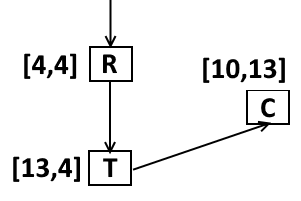}} \\

BlackEnergy \cite{black} &2013&2014 &Changes TESTSIGNING boot configuration option&Protective &Kernel level rootkit&[4,1]\\

IRATEMONK \cite{nsaant}&2008&2013 & MBR substitution & Access &Inappropriate General Access to HW/External control of Hw device & [11,4], [11,5]\\

DEITYBOUNCE \cite{nsaant}&2008&2013
& Modifies SMM code infection & Access&Inappropriate General Access to HW/External control of Hw device & [11,4], [11,5]\\

SWAP \cite{nsaant}&2008&2013
&Runs from hard-drive protected area&Sequential&Inappropriate General Access to HW/External control of Hw device&[10,4]\\

Cuttonmouth \cite{nsaant}&2008&2013&Spys data on the wire &Physical&Unexpected Consequences of Specific HW Function
&[13,13]\\

Sonic Screwdriver \cite{vault}&2012&2017&Thunderbolt Option ROM modifies EFI firmware&Sequential&HW Reflected Injection &\raisebox{-.6\height}{\includegraphics[scale=0.2]{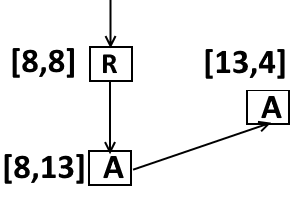}}\\

DarkSeaSkies \cite{vault}&2009&2017&Hides user-space implant&Logical &kernel rootkit&[4,1] \\

   \bottomrule[\heavyrulewidth] 
   \end{tabular}

\end{sidewaysfigure}

 Despite, we have gathered disclosed platform attacks and filtered those which works on a system with mentioned specification in section \ref{case}. We have also excluded those who solely rely on vulnerable implementation of some system component. Table \ref{tableeval} shows how the possibility of these attacks is provable by relations and metrics defined by Lamellae framework.

As the table suggests, platform attacks often occur in the context of nation-state cyber espionage and thus often remain undiscovered for years. Hence having the ability to identify chinks in the system's armour would enable us to forecast the possibility of such attacks.

\section{Discussion and future works}
\label{discuss}
Architectural attacks cannot be covered neither by hardware, nor by software security deciplines which analyze the SW/HW separately. Hence it is important to have architecture analysis routines against such attacks. Despite, there are multiple points about Lamellae to be considered.

First, Lamellae is intended to identify vulnerable architectural weak points exist in each platform by design.  Although applying changes in the platform's design is not possible by the end-user party, the framework enables the security analyzer to locate required security solutions/monitors at appropriate points of the system. It would be also possible to compare platforms based on their architectural security features.

Lamellae basic use-case is to model architectural threats against computing platforms. Despite when considering the existance of System-on-Chips (SoC) inside a platfrom, repeating the method recursively can be used also for finding some micro-architectural attacks. 
In fact, Lamellae is able to reveal attacks which exploits inter-relationship between components (through modules, SoC, FPGASoC, and etc.). Hence attacks which targets physical features of a single components (.e.g. Rowhammer \cite{rowhammer}), or micro-architectural implementation of an ISA code (.e.g. Out-of-Order Execution \cite{meltdown} attacks) are not in the scope of Lamellae. 

Another important point is that the SSA of a given platform which is made by Lamellae, can only represent the available information (by official documentation, reverse engineering efforts or etc.) about the platforms component. Hence it is no guarantee for the matrix to be complete. The origin of this problem is  in current abstract HW-SW contract which is limited to the processor ISA. Recently there have been arguments \cite{heiser}\cite{lowe} about insufficiently of the ISA and the need to more exposing contract.

One may argue that DSM analysis would not be easy when having a platform with many components. In such cases one can use tools exist to support DSM analysis (.e.g. Lattix\footnote{http://lattix.com/dependency-structure-matrix} and NDepends\footnote{https://www.ndepend.com/docs/dependency-structure-matrix-dsm}). This would be applicable when requiring to conform a more fine-grained analysis in a recursive manner for a platform.

Finally, Lamellae requires ad-hoc procedure for extracting initial set of code execution units of the platform. While we have suggested best practices to help doing this phase, we are working on a comprehensive ontology which can better guide users.

\section{conclusion}
\label{conclusion}
In this paper, we have discussed the lack of a threat modeling scheme which is applicable for considering architectural attacks against a computing platform. 

 We have proposed Lamellae, a platform threat modeling framework, which leverages privileged relations between platform components in order to achieve a multi-layered model of the system. Using the Design Structure Matrix (DSM) as the basic structure, we have altered the DFD representation by SSA-based DSM of the platform. We also used our taxonomy of original relations exploited during a platform attack, DSM-based analysis, and matrix-based metrics in order to detect threats against the platform.
  
Borrowing the threat modeling concept from the software security and the DSM analysis from the system's engineering literature, we have tried to adapt the existing analysis ideas to be applicable for finding architecturally weak points in a given computing platform. At the end, we have analyzed the security of an X86-64 architecture as a case-study.

\section{ APPENDIX A: BEST PRACTICES FOR IDENTIFYING PLATFORM COMPONENTS }
\begin{itemize}
   \item Start with the main board of the system, you can find a general block diagram of hardware components at the data-sheet/manual of the board. Identify components and their available interfaces from the manual.
   \item Do a further search about IC numbers you have found in the previous step. Look for the available information about their probable processor/controller and their capabilities.
    \item Boot sequence is a golden key word which helps you to find available data about hardware and firmware components and the sequential privileges in your platform. Even if the boot sequence was not mentioned from the manufacturer parties, you may find some information from reverse engineering results on the web.
   \item Enumerate CPU operational modes and different codes available for the modes.
   \item Enumerate all kinds of privileged codes can run on your CPU. This may include OS, hypervisor, interrupts, etc.
    \item Enumerate all hardware extensions supported by your specific CPU, software supports for the extensions, and available security reviews and attacks for them.
    \item Try to classify found components under user-level software, system software, firmware, and hardware categories.
\end{itemize}
\bibliographystyle{alpha}      
\bibliography{ref_related}

\newcommand{\etalchar}[1]{$^{#1}$}
\begin{thebibliography}{LRGCH18}

\bibitem[ACYA17]{almohri}
Hussain Almohri, Long Cheng, Danfeng Yao, and Homa Alemzadeh.
\newblock On threat modeling and mitigation of medical cyber-physical systems.
\newblock In {\em Connected Health: Applications, Systems and Engineering
  Technologies (CHASE), 2017 IEEE/ACM International Conference on}, pages
  114--119. IEEE, 2017.

\bibitem[AD10]{dma-windows}
Damien Aumaitre and Christophe Devine.
\newblock Subverting windows 7 x64 kernel with dma attacks.
\newblock {\em HITBSecConf Amsterdam}, 2010.

\bibitem[AGK15]{telehealth}
Mohamed Abomhara, Martin Gerdes, and Geir~M K{\o}ien.
\newblock A stride-based threat model for telehealth systems.
\newblock {\em Norsk informasjonssikkerhetskonferanse (NISK)}, 8(1):82--96,
  2015.

\bibitem[ALVW10]{alazab}
Manoun Alazab, Robert Layton, Sitalakshmi Venkataraman, and Paul Watters.
\newblock Malware detection based on structural and behavioural features of api
  calls.
\newblock In {\em 1st International Cyber Resilience Conference}. School of
  Computer and Information Science, Security Research Centre, Edith Cowan
  University, Perth, Western Australia, 2010.

\bibitem[ANW{\etalchar{+}}10]{hypersentry}
Ahmed~M Azab, Peng Ning, Zhi Wang, Xuxian Jiang, Xiaolan Zhang, and Nathan~C
  Skalsky.
\newblock Hypersentry: enabling stealthy in-context measurement of hypervisor
  integrity.
\newblock In {\em Proceedings of the 17th ACM conference on Computer and
  communications security}, pages 38--49. ACM, 2010.

\bibitem[AWJS16]{tmcloud1}
Nawaf Alhebaishi, Lingyu Wang, Sushil Jajodia, and Anoop Singhal.
\newblock Threat modeling for cloud data center infrastructures.
\newblock In {\em International Symposium on Foundations and Practice of
  Security}, pages 302--319. Springer, 2016.

\bibitem[BD12]{brossard2012hardware}
Jonathan Brossard and Florentin Demetrescu.
\newblock Hardware backdooring is practical.
\newblock {\em BlackHat, Las Vegas, USA}, 2012.

\bibitem[BDPV15]{gpu-assisted}
Davide Balzarotti, Roberto Di~Pietro, and Antonio Villani.
\newblock The impact of gpu-assisted malware on memory forensics: A case study.
\newblock {\em Digital Investigation}, 14:S16--S24, 2015.

\bibitem[Bil09]{arsenal}
Blunden Bill.
\newblock The rootkit arsenal. escape and evasion in the dark corners of the
  system, 2009.

\bibitem[bla14]{black}
Blackenergy and quedagh, 2014.

\bibitem[Blu12]{kernelrootkit2}
Bill Blunden.
\newblock {\em The Rootkit arsenal: Escape and evasion in the dark corners of
  the system}.
\newblock Jones \& Bartlett Publishers, 2012.

\bibitem[Bro16]{browning2016design}
Tyson~R Browning.
\newblock Design structure matrix extensions and innovations: a survey and new
  opportunities.
\newblock {\em IEEE Transactions on Engineering Management}, 63(1):27--52,
  2016.

\bibitem[BRPB14]{mal-cpu2}
Georg~T Becker, Francesco Regazzoni, Christof Paar, and Wayne~P Burleson.
\newblock Stealthy dopant-level hardware trojans: extended version.
\newblock {\em Journal of Cryptographic Engineering}, 4(1):19--31, 2014.

\bibitem[BSD08]{biossmm}
Coideloko BSDaemon.
\newblock D0nand0n. system management mode hack: Using smm for" other purposes.
\newblock {\em Phrack Magazine}, 12:65, 2008.

\bibitem[BSS10]{bahmani2010survey}
Faezeh Bahmani, Marzieh Shariati, and Fereidoon Shams.
\newblock A survey of interoperability in enterprise information security
  architecture frameworks.
\newblock In {\em Information Science and Engineering (ICISE), 2010 2nd
  International Conference on}, pages 1794--1797. IEEE, 2010.

\bibitem[CA14]{microcode}
Daming~D Chen and Gail-Joon Ahn.
\newblock Security analysis of x86 processor microcode, 2014.

\bibitem[CKR16]{ccaf2}
Victor Chang, Yen-Hung Kuo, and Muthu Ramachandran.
\newblock Cloud computing adoption framework: A security framework for business
  clouds.
\newblock {\em Future Generation Computer Systems}, 57:24--41, 2016.

\bibitem[CR16]{ccaf1}
Victor Chang and Muthu Ramachandran.
\newblock Towards achieving data security with the cloud computing adoption
  framework.
\newblock {\em IEEE Transactions on Services Computing}, 9(1):138--151, 2016.

\bibitem[DBL09]{exposed}
Michael Davis, Sean Bodmer, and Aaron LeMasters.
\newblock {\em Hacking Exposed Malware and Rootkits}.
\newblock McGraw-Hill, Inc., 2009.

\bibitem[DHWW11]{cloak-comp}
Alan~M Dunn, Owen~S Hofmann, Brent Waters, and Emmett Witchel.
\newblock Cloaking malware with the trusted platform module.
\newblock In {\em USENIX Security Symposium}, 2011.

\bibitem[Dom13]{mal-disk}
Jeroen Domburg.
\newblock Hard disk hacking.
\newblock {\em http://spritesmods.com}, 2013.

\bibitem[EB12]{dsm-methods}
Steven~D Eppinger and Tyson~R Browning.
\newblock {\em Design structure matrix methods and applications}.
\newblock MIT press, 2012.

\bibitem[EMO12]{elhadi}
Ammar Ahmed~E Elhadi, Mohd~Aizaini Maarof, and Ahmed~Hamza Osman.
\newblock Malware detection based on hybrid signature behaviour application
  programming interface call graph.
\newblock {\em American Journal of Applied Sciences}, 9(3):283, 2012.

\bibitem[ES08]{embleton}
Shawn Embleton and Sherri Sparks.
\newblock Smm rootkits.
\newblock In {\em Proceedings of the 4th International Conference on Security
  and Privacy in Communication Netowrks, Securecomm}, volume~8, 2008.

\bibitem[ESZ13]{smmrootkit}
Shawn Embleton, Sherri Sparks, and Cliff~C Zou.
\newblock Smm rootkit: a new breed of os independent malware.
\newblock {\em Security and Communication Networks}, 6(12):1590--1605, 2013.

\bibitem[For11]{forristal}
Jeff Forristal.
\newblock Hardware involved software attacks.
\newblock {\em CanSecWest, Vancouver, Canada}, 2011.

\bibitem[GD09]{lala}
Carl Gebhardt and Chris Dalton.
\newblock Lala: a late launch application.
\newblock In {\em Proceedings of the 2009 ACM workshop on Scalable trusted
  computing}, pages 1--8. ACM, 2009.

\bibitem[GE{\etalchar{+}}91]{circuits}
David~A Gebala, Steven~Daniel Eppinger, et~al.
\newblock {\em Methods for analyzing design procedures}.
\newblock Sloan School of Management, Massachusetts Institute of Technology,
  1991.

\bibitem[Han05]{SSA}
Susan Hansche.
\newblock {\em Official (ISC) 2{\textregistered} Guide to the
  CISSP{\textregistered}-ISSEP{\textregistered} CBK{\textregistered}}.
\newblock CRC Press, 2005.

\bibitem[Hay02]{bus-hierarchies2}
John~P Hayes.
\newblock {\em Computer architecture and organization}.
\newblock McGraw-Hill, Inc., 2002.

\bibitem[Hea06]{acpi-rootkit}
John Heasman.
\newblock Implementing and detecting an acpi bios rootkit.
\newblock {\em Black Hat Federal}, 368, 2006.

\bibitem[Hei17]{heiser}
Gernot Heiser.
\newblock For safety’s sake: We need a new hardware-software contract!
\newblock {\em IEEE Design \& Test}, 2017.

\bibitem[HTLC15]{killchain}
Adam Hahn, Roshan~K Thomas, Ivan Lozano, and Alvaro Cardenas.
\newblock A multi-layered and kill-chain based security analysis framework for
  cyber-physical systems.
\newblock {\em International Journal of Critical Infrastructure Protection},
  11:39--50, 2015.

\bibitem[Hu10]{hids2}
Jiankun Hu.
\newblock Host-based anomaly intrusion detection.
\newblock In {\em Handbook of Information and Communication Security}, pages
  235--255. Springer, 2010.

\bibitem[Joh13]{optionROM}
Saul~St John.
\newblock Thunderbolt: Exposure and mitigation, 2013.

\bibitem[KCKB14]{suppression}
C~Kallenberg, C~Cornwell, X~Kovah, and J~Butterworth.
\newblock Setup for failure: More ways to defeat secureboot.
\newblock {\em Hack In The Box Amsterdam, Amsterdam}, 2014.

\bibitem[KGG{\etalchar{+}}18]{spe}
Paul Kocher, Daniel Genkin, Daniel Gruss, Werner Haas, Mike Hamburg, Moritz
  Lipp, Stefan Mangard, Thomas Prescher, Michael Schwarz, and Yuval Yarom.
\newblock Spectre attacks: Exploiting speculative execution.
\newblock {\em ArXiv e-prints}, 2018.

\bibitem[KJM{\etalchar{+}}11]{trustcloud}
Ryan~KL Ko, Peter Jagadpramana, Miranda Mowbray, Siani Pearson, Markus
  Kirchberg, Qianhui Liang, and Bu~Sung Lee.
\newblock Trustcloud: A framework for accountability and trust in cloud
  computing.
\newblock In {\em 2011 IEEE World Congress on Services}, pages 584--588. IEEE,
  2011.

\bibitem[KK15]{lighteater}
Corey Kallenberg and Xeno Kovah.
\newblock How many million bioses would you like to infect?, 2015.

\bibitem[Kle09]{stonedbootkit}
Peter Kleissner.
\newblock Stoned bootkit.
\newblock In {\em Black Hat USA}, 2009.

\bibitem[Kro12]{perspectives}
John Krogstie.
\newblock Perspectives to process modeling--a historical overview.
\newblock In {\em Enterprise, Business-Process and Information Systems
  Modeling}, pages 315--330. Springer, 2012.

\bibitem[KTC{\etalchar{+}}08]{mal-cpu}
Samuel~T King, Joseph Tucek, Anthony Cozzie, Chris Grier, Weihang Jiang, and
  Yuanyuan Zhou.
\newblock Designing and implementing malicious hardware.
\newblock {\em LEET}, 8:1--8, 2008.

\bibitem[LCH{\etalchar{+}}16]{speccert}
Thomas Letan, Pierre Chifflier, Guillaume Hiet, Pierre N{\'e}ron, and Benjamin
  Morin.
\newblock Speccert: specifying and verifying hardware-based security
  enforcement.
\newblock In {\em International Symposium on Formal Methods}, pages 496--512.
  Springer, 2016.

\bibitem[LKV{\etalchar{+}}13]{dmakeylogger}
Evangelos Ladakis, Lazaros Koromilas, Giorgos Vasiliadis, Michalis
  Polychronakis, and Sotiris Ioannidis.
\newblock You can type, but you can’t hide: A stealthy gpu-based keylogger.
\newblock In {\em Proceedings of the 6th European Workshop on System Security
  (EuroSec)}, 2013.

\bibitem[LLZ{\etalchar{+}}13]{mguard}
Ziyi Liu, JongHyuk Lee, Junyuan Zeng, Yuanfeng Wen, Zhiqiang Lin, and Weidong
  Shi.
\newblock {\em Cpu transparent protection of os kernel and hypervisor integrity
  with programmable dram}, volume~41.
\newblock ACM, 2013.

\bibitem[LPAF{\etalchar{+}}18]{lowe}
Jason Lowe-Power, Venkatesh Akella, Matthew~K Farrens, Samuel~T King, and
  Christopher~J Nitta.
\newblock A case for exposing extra-architectural state in the isa: position
  paper.
\newblock In {\em Proceedings of the 7th International Workshop on Hardware and
  Architectural Support for Security and Privacy}, page~8. ACM, 2018.

\bibitem[LRGCH18]{freespec}
Thomas Letan, Yann R{\'e}gis-Gianas, Pierre Chifflier, and Guillaume Hiet.
\newblock Modular verification of programs with effects and effect handlers in
  coq.
\newblock In {\em FM 2018-22nd International Symposium on Formal Methods},
  2018.

\bibitem[LSG{\etalchar{+}}18]{meltdown}
Moritz Lipp, Michael Schwarz, Daniel Gruss, Thomas Prescher, Werner Haas,
  Stefan Mangard, Paul Kocher, Daniel Genkin, Yuval Yarom, and Mike Hamburg.
\newblock Meltdown.
\newblock {\em ArXiv e-prints}, 2018.

\bibitem[LV03]{lawlor2003survey}
Brendan Lawlor and Linh Vu.
\newblock A survey of techniques for security architecture analysis.
\newblock Technical report, DEFENCE SCIENCE AND TECHNOLOGY ORGANISATION
  SALISBURY (AUSTRALIA) INFO SCIENCES LAB, 2003.

\bibitem[LZO10]{hids1}
Ying Lin, Yan Zhang, and Yang-jia Ou.
\newblock The design and implementation of host-based intrusion detection
  system.
\newblock In {\em Intelligent Information Technology and Security Informatics
  (IITSI), 2010 Third International Symposium on}, pages 595--598. IEEE, 2010.

\bibitem[MBK{\etalchar{+}}15]{tm-cps}
Goncalo Martins, Sajal Bhatia, Xenofon Koutsoukos, Keith Stouffer, CheeYee
  Tang, and Richard Candell.
\newblock Towards a systematic threat modeling approach for cyber-physical
  systems.
\newblock In {\em Resilience Week (RWS), 2015}, pages 1--6. IEEE, 2015.

\bibitem[Met10]{mcr}
Erez Metula.
\newblock {\em Managed code rootkits: hooking into runtime environments}.
\newblock Elsevier, 2010.

\bibitem[MPP{\etalchar{+}}08]{flicker}
Jonathan~M McCune, Bryan~J Parno, Adrian Perrig, Michael~K Reiter, and Hiroshi
  Isozaki.
\newblock Flicker: An execution infrastructure for tcb minimization.
\newblock In {\em ACM SIGOPS Operating Systems Review}, volume~42, pages
  315--328. ACM, 2008.

\bibitem[MR11]{alureon}
Aleksandr Matrosov and Eugene Rodionov.
\newblock Defeating x64: Modern trends of kernel-mode rootkits.
\newblock In {\em Virusbulletin}, 2011.

\bibitem[nsa13]{nsaant}
Nsa ant catalog, 2013.

\bibitem[Pel04]{kernelrootkit1}
R{\'a}ul~Siles Pel{\'a}ez.
\newblock Linux kernel rootkits: protecting the system’s” ring-zero”.
\newblock {\em GIAC Unix Security Administrator (GCUX)}, page 169, 2004.

\bibitem[PSM{\etalchar{+}}13]{cyberconflict}
K~Podins, J~Stinissen, M~Maybaum, et~al.
\newblock Towards a cyber conflict taxonomy.
\newblock In {\em 5th International Conference on Cyber Conflict}. Citeseer,
  2013.

\bibitem[RAJ16]{cross-layer}
JYRI RAJAM{\"A}KI.
\newblock Cross-layer approach for designing resilient (sociotechnical,
  cyber-physical, software-intensive and systems of) systems.
\newblock {\em International Journal of Communications}, pages 137--144, 2016.

\bibitem[RT08]{bluepill}
Joanna Rutkowska and Alexander Tereshkin.
\newblock Bluepilling the xen hypervisor.
\newblock {\em Black Hat USA}, 2008.

\bibitem[Rua14]{MEbook}
Xiaoyu Ruan.
\newblock {\em Platform Embedded Security Technology Revealed: Safeguarding the
  Future of Computing with Intel Embedded Security and Management Engine}.
\newblock Apress, 2014.

\bibitem[SB12]{dagger}
Patrick Stewin and Iurii Bystrov.
\newblock Understanding dma malware.
\newblock In {\em International Conference on Detection of Intrusions and
  Malware, and Vulnerability Assessment}, pages 21--41. Springer, 2012.

\bibitem[SB13]{ME2013}
Patrick Stewin and Iurii Bystrov.
\newblock Persistent, stealthy, remote-controlled dedicated hardware malware.
\newblock In {\em Chaos Communication Congress. Berlin}, 2013.

\bibitem[SC13]{dma-pci}
Johannes St{\"u}ttgen and Michael Cohen.
\newblock Anti-forensic resilient memory acquisition.
\newblock {\em Digital Investigation}, 10:S105--S115, 2013.

\bibitem[SCK{\etalchar{+}}16]{pikit}
Wonjun Song, Hyunwoo Choi, Junhong Kim, Eunsoo Kim, Yongdae Kim, and John Kim.
\newblock Pikit: A new kernel-independent processor-interconnect rootkit.
\newblock In {\em USENIX Security Symposium}, pages 37--51, 2016.

\bibitem[SCL95]{sherwood}
John Sherwood, Andrew Clark, and David Lynas.
\newblock Enterprise security architecture.
\newblock {\em SABSA, White paper}, 2009, 1995.

\bibitem[SES{\etalchar{+}}09]{avoidit}
Chris Simmons, Charles Ellis, Sajjan Shiva, Dipankar Dasgupta, and Qishi Wu.
\newblock Avoidit: A cyber attack taxonomy, 2009.

\bibitem[She05]{sherwood2005enterprise}
Nicholas~A Sherwood.
\newblock {\em Enterprise security architecture: a business-driven approach}.
\newblock CRC Press, 2005.

\bibitem[SHL16]{lophi}
Chad Spensky, Hongyi Hu, and Kevin Leach.
\newblock Lo-phi: Low-observable physical host instrumentation for malware
  analysis.
\newblock In {\em Proceedings of the Network and Distributed System Security
  Symposium}, 2016.

\bibitem[Sho14]{threatmodelbook}
Adam Shostack.
\newblock {\em Threat modeling: Designing for security}.
\newblock John Wiley \& Sons, 2014.

\bibitem[Ska07]{v-black}
Kevin Skapinetz.
\newblock Virtualisation as a blackhat tool.
\newblock {\em Network Security}, 2007(10):4--7, 2007.

\bibitem[SLGM12]{Shosha2012}
Ahmed~F. Shosha, Chen~Ching Liu, Pavel Gladyshev, and Marcus Matten.
\newblock {Evasion-resistant malware signature based on profiling kernel data
  structure objects}.
\newblock {\em 7th International Conference on Risks and Security of Internet
  and Systems, CRiSIS 2012}, 2012.

\bibitem[Sta00]{bus-hierarchies}
William Stallings.
\newblock {\em Computer organization and architecture: designing for
  performance}.
\newblock Pearson Education India, 2000.

\bibitem[Ste11]{ME2011}
Patrick Stewin.
\newblock Evaluating ring -3 rootkits, 2011.

\bibitem[stm15]{stm-userguid}
Smi transfer monitor (stm)-user guide, 2015.

\bibitem[Tor05]{torr2005demystifying}
Peter Torr.
\newblock Demystifying the threat modeling process.
\newblock {\em IEEE Security \& Privacy}, 3(5):66--70, 2005.

\bibitem[TW09]{ME2009}
Alexander Tereshkin and Rafal Wojtczuk.
\newblock Introducing ring -3 rootkits.
\newblock In {\em Blackhat US}, 2009.

\bibitem[vau17]{vault}
Vault 7: Cia hacking tools revealed, 2017.

\bibitem[Win17]{winsen2017threat}
Stijn Winsen.
\newblock Threat modelling for future vehicles: on identifying and analysing
  threats for future autonomous and connected vehicles.
\newblock Master's thesis, University of Twente, 2017.

\bibitem[WR09a]{invisible}
Rafal Wojtczuk and Joanna Rutkowska.
\newblock Attacking intel trusted execution technology.
\newblock {\em Blackhat DC 2009}, 2009.

\bibitem[WR09b]{cachep}
Rafal Wojtczuk and Joanna Rutkowska.
\newblock Attacking smm memory via intel cpu cache poisoning, 2009.

\bibitem[WR11]{against-vt}
Rafal Wojtczuk and Joanna Rutkowska.
\newblock Following the white rabbit: Software attacks against intel vt-d
  technology, 2011.

\bibitem[WSG10]{hypercheck}
Jiang Wang, Angelos Stavrou, and Anup Ghosh.
\newblock {HyperCheck: A hardware-assisted integrity monitor}.
\newblock In {\em Recent Advances in Intrusion Detection}, pages 158--177.
  Springer, 2010.

\bibitem[YHK{\etalchar{+}}12]{dfd}
Mark Yampolskiy, Peter Horvath, Xenofon~D Koutsoukos, Yuan Xue, and Janos
  Sztipanovits.
\newblock Systematic analysis of cyber-attacks on cps-evaluating applicability
  of dfd-based approach.
\newblock In {\em Resilient Control Systems (ISRCS), 2012 5th International
  Symposium on}, pages 55--62. IEEE, 2012.

\bibitem[ZSL{\etalchar{+}}15]{see}
Ning Zhang, Kun Sun, Wenjing Lou, Y~Thomas Hou, and Sushil Jajodia.
\newblock Now you see me: Hide and seek in physical address space.
\newblock In {\em Proceedings of the 10th ACM Symposium on Information,
  Computer and Communications Security}, pages 321--331. ACM, 2015.

\bibitem[ZWLS14]{iocheck}
Fengwei Zhang, Haining Wang, Kevin Leach, and Angelos Stavrou.
\newblock A framework to secure peripherals at runtime.
\newblock In {\em Computer Security-ESORICS 2014}, pages 219--238. Springer,
  2014.

\bibitem[ZZS{\etalchar{+}}18]{zhang}
Ning Zhang, Ruide Zhang, Kun Sun, Wenjing Lou, Y~Thomas Hou, and Sushil
  Jajodia.
\newblock Memory forensic challenges under misused architectural features.
\newblock {\em IEEE Transactions on Information Forensics and Security},
  13(9):2345--2358, 2018.

\end{thebibliography}

\end{document}